\documentclass[12pt]{article}
\usepackage{graphicx}
\usepackage{comment}
\usepackage{amssymb}
\usepackage[a4paper, total={7in, 8in}]{geometry}

\usepackage{amstext}
\usepackage{amsmath}
\usepackage{amsthm}
\usepackage{bm} % for bold stuff in math environment
\usepackage{commath}
\usepackage{float}
\usepackage{graphicx}
\usepackage{mathtools}
\usepackage{hyperref}
\usepackage{xcolor}
\usepackage{xargs}  
\usepackage{subfigure}
\begin{document}
%-------

%-------
\title{Some exactly solvable and tunable frustrated spin models}
\author{F. Caravelli\\Theoretical Division (T4),\\
Los Alamos National Laboratory,\\ Los Alamos, New Mexico 87545, USA}
%\ead{caravelli@lanl.gov}
\vspace{10pt}
%\begin{indented}
%\item[]August 2017
%\end{indented}
\maketitle

\begin{abstract}
We discuss three exactly solvable spin  models of geometric frustration. First, we discuss a 1-parameter subfamily of the 16 vertex model, which can be mapped to a planar Ising model and solved via Fisher-Dubed\'{a}t decorations. We then consider a 1-parameter family generalization of the Villain's fully frustrated model, which interpolates between Onsager's 2D Ising model and the Villain one. We then discuss spin ice models on a tree, which can be solved exactly using recursions \textit{a l\'{a} Bethe}. 
\end{abstract}

%\maketitle

\section{Introduction}

Over the last decade there has been a renewed interest in the study of ice models. Such interest is due to the fact that spin ice materials can be engineered using nanomagnets, so called Artificial Spin Ices \cite{colloq,Wang1,Bader,reddim,Heyderman,Canals1,Nisoli1,Morgan,Budrikis,Branford}. In artificial spin ice, which are particular metamaterials, the low energy models are those of classical spin ices, while it is typically harder to engineer the energetics of the higher order excitations.
It is however becoming increasingly possible to choose the energy hierarchies of the vertices \cite{Nisoli4,mol,Castelnovo,Castelnovo2, topor,Cugliandolo2,logic5,gartside,WangYL2,Gilbert,Nisoli8}, by placing either dot-islands at the center of each vertex,  carefully choosing the relative heights of the islands \cite{modifiers,caravelli}, or via changing the height of the islands \cite{ladak,Faran1}. 

In view of such interest, it is worth re-examining ice models \cite{Lieb3} with some old tools  and new methods. 
While the Rys F-Model \cite{Lieb,Nisolirys}, and the 6 and 8 vertex models can be exactly solved \cite{Lieb2,Baxter}, the 16 vertex model is particularly hard to solve in its full generality. One of the key reasons is that the 16-vertex model is mapped to a non-planar Ising model, which cannot be solved via transfer method techniques or dimer mapping. A possible way of seeing this is by noticing that the effective Ising model into which the 16 vertex model can be mapped is non-planar, and as such non-integrable in its full generality. Using the well known transfer matrix methods, it possible to see that there are however some integrability conditions for the parameters in which case the 16 vertex model can be solved. The two main families of models, a subclass of the 16 vertex model, that can be solved are the so called even- and odd- 8 vertex models.

Similarly, Onsager's solution, first reobtained by Kac and Ward via a combinatorial method, had then been re-derived by Montroll, Potts and Ward (MPW) using, in fact, the solution of the dimer problem for a planar lattice. Given a certain planar graph, solving a dimer problem means finding the number of ways in which the edges of the graph, and thus pair of vertices, could be covered with non-overlapping dominos. The solution of such problem was obtained essentially in parallel by Kasteleyn and Fisher and Temperley.
While the Montroll-Potts-Ward (MPW) solution of the 2D Ising model was obtained by a magical cancellation, until transfer methods and the Bethe Ansatz became of common use and better understood, mapping an Ising model to a dimer problem often meant solving the model \cite{kasteleyn,tempfisher}. It has been known for long time that, in fact, all planar Ising models can be in principle be solved \cite{Fisher}.  Fisher was the first to realize that in fact any planar Ising model could be solved via the mapping to an equivalent dimer problem without magical cancellations. Such construction is often called Fisher decorations: vertices of the Ising model are mapped to larger planar graph structures composed of triangles, for which a Kasteleyn orientation (e.g. an orientation of the edges of the graph such that every cycle of the graph is odd) always exists \cite{montroll,Kenyon}.
The drawback of such technique is that for graphs with vertices of relatively large degree, and in particular at a time when computers and algebra software was not available, the vertex proliferation meant calculating large determinants. Nowadays, however, we have both advanced algebra software able to calculate large parametric determinants and new Fisher decorations for which the vertex proliferation is diminished compared to the Fisher proposal. 

In the present paper we proceed in a different way towards the derivation of a special but exact solution of 16 vertex model, which can be expressed in the identical form to the odd 8-vertex model. In this sense, the results of this paper should not come as a complete surprise to the specialists of the topic, but as far as we know the methodology we employ is, to some extent, new for this particular application.
We use a recent decoration suggested by Dub\'{e}dat, which we will elucidate in a moment, to map the planar Ising model to a dimer model on a decorated lattice. We then evaluate the determinant in order to solve for the partition function of the model.

We consider two other models in which we have a tunable frustration. The first is Villain's fully frustrated Ising model. Here we consider a slight generalization which interpolates between the Ising model and Villain's model, and which can be still be solved using standard dimer techniques. The third model we consider is effective square ice interaction model on a tree, which can be solved exactly using standard ``Bethe" equations, e.g. by integrating leaves of a tree out. Spin ice models on a tree have been recently investigated experimentally \cite{sacconetree} and thus this subject is of direct practical relevance.

Conclusions follow.

\section{A 1-parameter subfamily of the 16 vertex model via planar Ising models}

\begin{figure}[b!]
    \centering
    \includegraphics[scale=1.2]{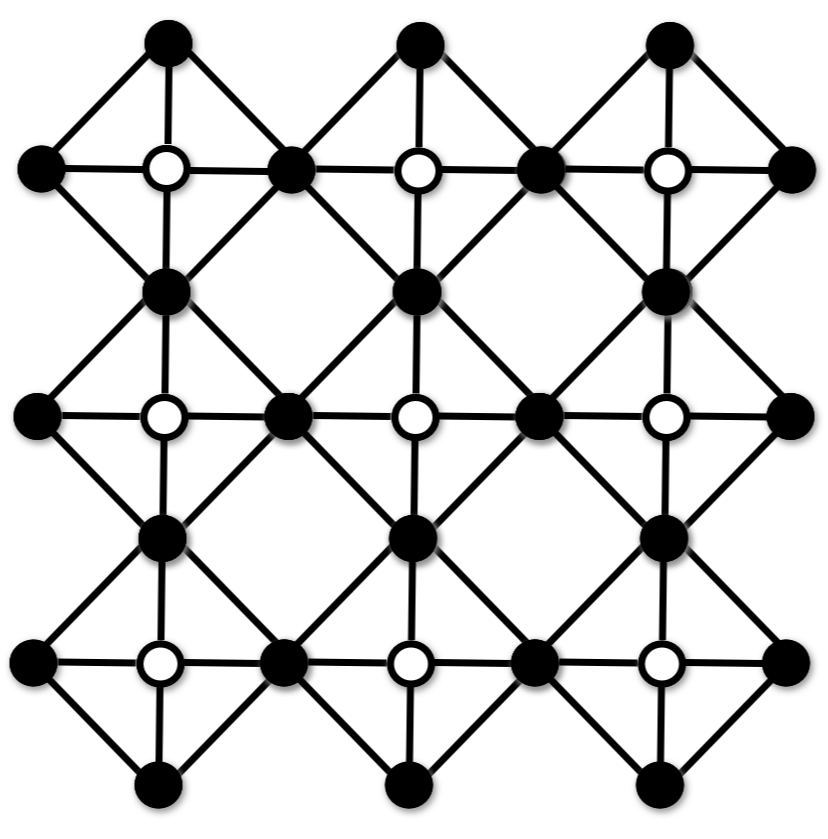}
    \caption{The staggered planar Ising model we consider in this paper. }
    \label{fig:lattice}
\end{figure}

\begin{figure*}[bt!]
    \centering
    \includegraphics[scale=1.5]{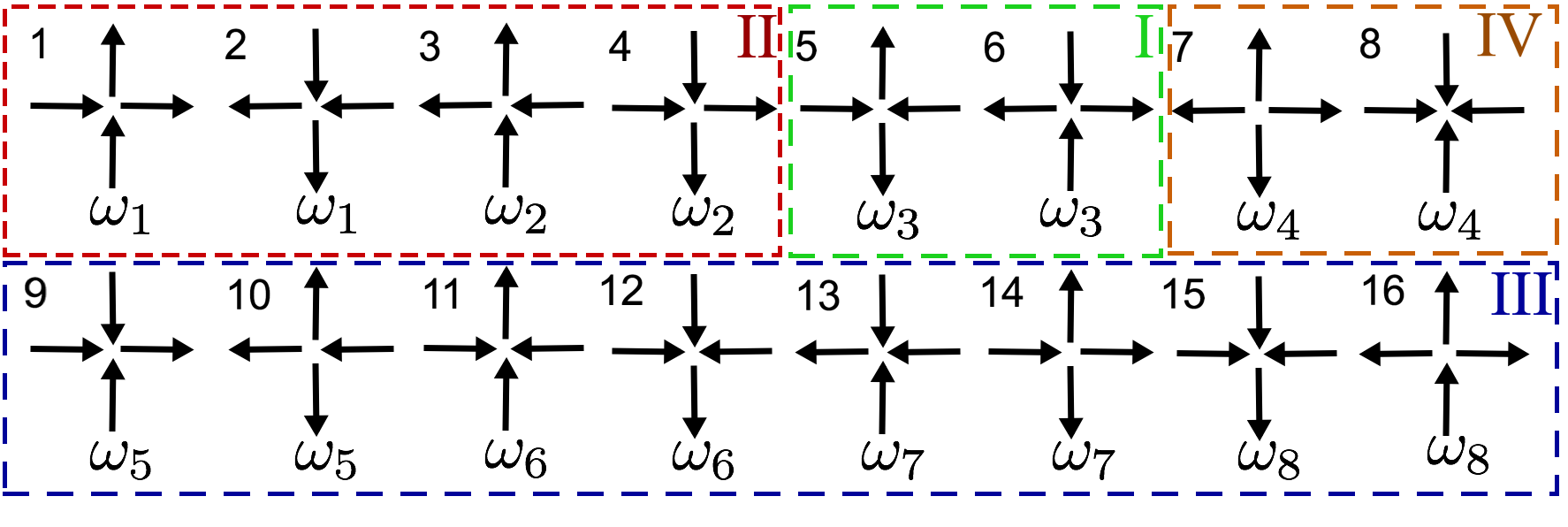}
    \caption{The sixteen vertex model and the classification in terms of vertices of Type I - Type IV in artificial spin ice.}
    \label{fig:vertexenergies}
\end{figure*}

\begin{figure}
    \centering
    \includegraphics[scale=2]{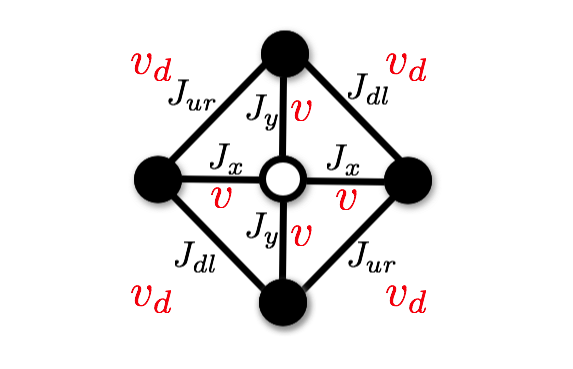}
    \caption{Parameters of the model (black) and their reduction after the vertex symmetry constraints (red). }
    \label{fig:couplings}
\end{figure}

The 16 vertex model \cite{Baxter,Wu1,Wu2,Wu3,baxter2,baxter3,Assis} is the model described a square lattice, whose vertex configurations can assume the configurations shown in Fig. \ref{fig:vertexenergies}. The energies of these vertices are typically labelled by $\omega_1=\cdots=\omega_{16}$. 
Our proposal is choose an effective planar Ising model which we can solve, in the spirit of Wu \cite{Wu1} and Wu and Lieb \cite{Lieb2}. The lattice is the one of Fig. (\ref{fig:lattice}), which represent a staggered Ising model, e.g. a model whose interactions change from plaquette to plaquette in a regular way. The black vertices represent the in-plane spins of the 16-vertex model, while the white vertices are auxiliary spins which modify locally the interaction between the vertices. In Wu's and Baxter's original proposals, for instance, the white vertex is not present and the interactions are directly between the black spins. However, because of the structure, this would imply that the model is not planar anymore. We parametrize the couplings of such Ising model as in Fig.  \ref{fig:couplings}. The vertex energies can then be obtained by inspecting the Ising energy for every vertex configuration. The result is the one of eqns. (\ref{eq:firsten})-(\ref{eq:lasten}):

\begin{eqnarray}
\label{eq:firsten}
    e^{\omega_1}&=& 2 e^{2 (J_{ur}+J_{dl})} \cosh(2 J_y+2 J_x) \\
    e^{\omega_2}&=&2 e^{-2 (J_{ur}+J_{dl})} \cosh(2 J_y-2 J_x)\\
    e^{\omega_3}&=&2 e^{2 (J_{ur}-J_{dl})} \\
    e^{\omega_4}&=&2 e^{-2 (J_{ur}-  J_{dl})}\\
    e^{\omega_5}&=& 2 \cosh(2 J_x)\\
    e^{\omega_6}&=& 2 e^{2(J_{ur}-J_{dl})}\cosh(2 J_y)\\
    e^{\omega_7}&=& 2 \cosh(2 J_x)\\
    e^{\omega_8}&=& 2\cosh(2 J_y)
    \label{eq:lasten}
\end{eqnarray}

It is not hard to see that in order for the Type I - Type IV division of the vertex energies, one has to have $J_x=J_y$ and $J_{ur}=J_{dl}$. Note that since the factor $2$ is there for every vertex, it can be removed via an energy shift (or equivalently via a partition function redefinition)  without affecting the thermodynamic properties.

It follows that if we want $\omega_5=\cdots=\omega_8$, we must have $J_x=J_y\equiv v$ and $J_{ur}=J_{ld}\equiv v_d$ (the ``d" stands for diagonal), and using this definition we have $\omega_5=\omega_6=\omega_7=\omega_8=\log \big(2 \cosh(2 v)\big)$. In this case, $\omega_1=  \log (e^{4 v_d} \cosh(4 v))$, $\omega_2= \log \big(2 e^{-4 v_d}\big)$, $\omega_3=\omega_4=\log 2$.

 The artificial square ice is thus described by the following Hamiltonian based on Ising-like variables, but lying on the plane:
\begin{equation}
    H_{ASI}=-\sum_{v}\Big[ \epsilon_{||}(\sum_{\langle i,j\rangle_v} s_{x}s_{x}+\sum_{\langle i,j\rangle_v} s^i_{y}s^j_{y} )+ \epsilon_{\perp} \sum_{\langle i,j\rangle_v} s^i_{x} s_{y} ^j\Big]
\end{equation}
where $\vec s_{x}=s_{x}\hat x$ and $\vec s_{y}=s_{y}\hat y$. 
For the square spin ice, it has been noted that vertices have four increasing energies parametrized by $\epsilon_{\perp}$ and $\epsilon_{||}$, with a nomenclature \textrm{Type I},$\cdots$,\textrm{Type IV} respectively. The vertex energies are
    $\epsilon_{I}=-4 \epsilon_{\perp}+2 \epsilon_{||}$,  $\epsilon_{II}=-2 \epsilon_{||}$, \\
    $\epsilon_{III}=0$,
    $\epsilon_{IV}= 4\epsilon_{\perp}+2\epsilon_{||}$,
where $\epsilon_{I}<\epsilon_{II}<\epsilon_{III}<\epsilon_{IV}$. The vertex population in the ground state is determined by this energy hierarchy. In units of the temperature when setting the Boltzmann constant $\kappa=1$, we can use 
$\epsilon_{\perp}\approx0.38675$ and $\epsilon_{||}\approx 0.2735$ for realistic phase diagrams, as noted in \cite{Morrison}.

As a result, we obtain the following mapping between the vertex energies and the spin ice energies:
\begin{eqnarray}
    e^{\epsilon_I}&=&e^{-4 \epsilon_{\perp}+2 \epsilon_{\|}}= 1\\
    e^{ \epsilon_{II}}&=&e^{-2 \epsilon_{\|}}= e^{\omega_1}=e^{\omega_2}= e^{-2 v_d} \\
   e^{ \epsilon_{III}}&=&e^{\omega_5}=\cdots =e^{\omega_8}=\cosh(2 v)\\
   e^{\epsilon_{IV}}&=&e^{4 \epsilon_{\perp}+2 \epsilon_{\|}}=1.
\end{eqnarray}

\begin{figure}
\centering
    \includegraphics[scale=1.2]{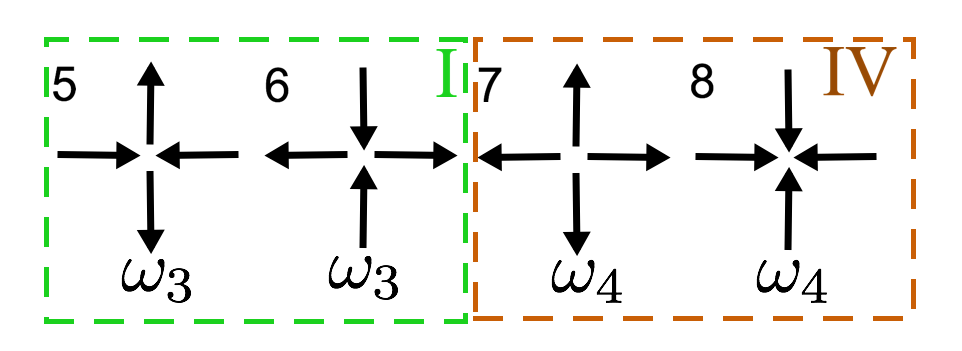}
    \caption{The vertices of the monopole gas for $v\gg 1$, with energies $\omega_3=\omega_4=0$.}
    \label{eq:mongas}
\end{figure}

It follows that the planarity requirement is too strong and the standard energetic parametrization for artificial spin ice cannot be described by the model we solved in this paper for arbitrary values of $v$ and $v_d$.  This said, there are a few limits of this model that are interesting.
This is not the case however for an interesting subcase of this model can be however of interest in artificial spin ice, and which is a subcase of the 8-vertex model.
 If we require that $\omega_1=\omega_2$, then we must have also the condition $e^{-8 v_d}=\cosh(4 v) $.  If we define the energies of the spin ice types, we have
\begin{eqnarray}
    \epsilon_I&=&0\\
    \epsilon_{II}&=&2 \log\cosh(4 v)\\
    \epsilon_{III}&=&\log \cosh(2 v)\\
    \epsilon_{IV}&=&0,
\end{eqnarray}
which is a particular 1-parameter subfamily of the vertex models typically of interest for artificial spin ice. It is interesting to note that in the limit $v\rightarrow \infty$, vertices of type II and III have infinite energies, and thus the model reduces to a particular monopole gas, with the two sets of vertices shown in Fig. \ref{eq:mongas}.
This is a 4-vertex model which incorporates two 0-charge vertices and two 4-charge vertices. It would be interesting if such model could be realized in practice, but in current experiments of artificial spin ice, typically the vertices of Type IV are always of higher energy.

\subsection{Fisher-Dub\'{e}dat decorations}

\begin{figure}
    \centering
    \includegraphics[scale=1.4]{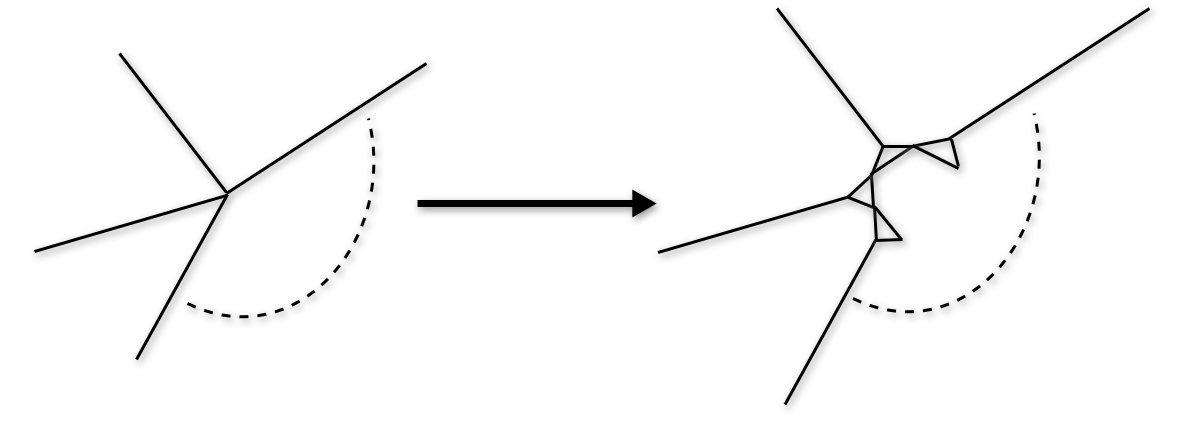}
    \caption{Decorated lattice using the Dubed\'{a}t transformation.}
    \label{fig:dubedat}
\end{figure}

Let us now discuss the method we employ to solve the model. As mentioned earlier, Fisher decorations can be applied to planar Ising models. In this paper we use variant introduced by Dubed\'{a}t \cite{detilier1,detilier2,dubedat}. Each Ising model vertex is converted in a series of triangles, as in Fig. \ref{fig:dubedat}. If the vertex has degree $D>2$, the number of triangles is exactly $D$. The number of vertices being introduced in the decorated lattice is $2D$. As it is the case both for the MPW and Fisher solutions, also in this case the dimer mapping is obtained via a high temperature expansion. Let $G$ be the graph representing the interactions of the planar Ising model and $G^{FD}$ the decorated lattice. Using the Fisher-Dub\'{e}dat decoration, we can identify edges that were in the graph $G$ and extra edges introduced in the decoration. Clearly, the edges $e$ of the original Ising model are weighted by the interactions $J_{e}$ on that specific edge. The partition function of the equivalent dimer problem is weighted by $\nu_e$.
Then
\begin{equation}
    \nu_e=\begin{cases}
    1 & \text{if the edge $e$ belongs to the decoration}\\
    \tanh(J_e) & \text{if the edge $e$ is in the original graph}\\
    0 & \text{otherwise}
    \end{cases}
\end{equation}
Since the Fisher-Dub\'{e}dat decoration is always planar, a Kasteleyn orientation can always be found. Given this, the mapping between the dimer partition function with oriented weights $\nu_e$ and the Ising model is simply given by
$Z_{Ising}(G,J)=\prod_{e\in E} \cosh(J_e) Z_{dimer}(G^{FD},\{\nu_e\})$.

\begin{figure}
    \centering
    \includegraphics[scale=1.2]{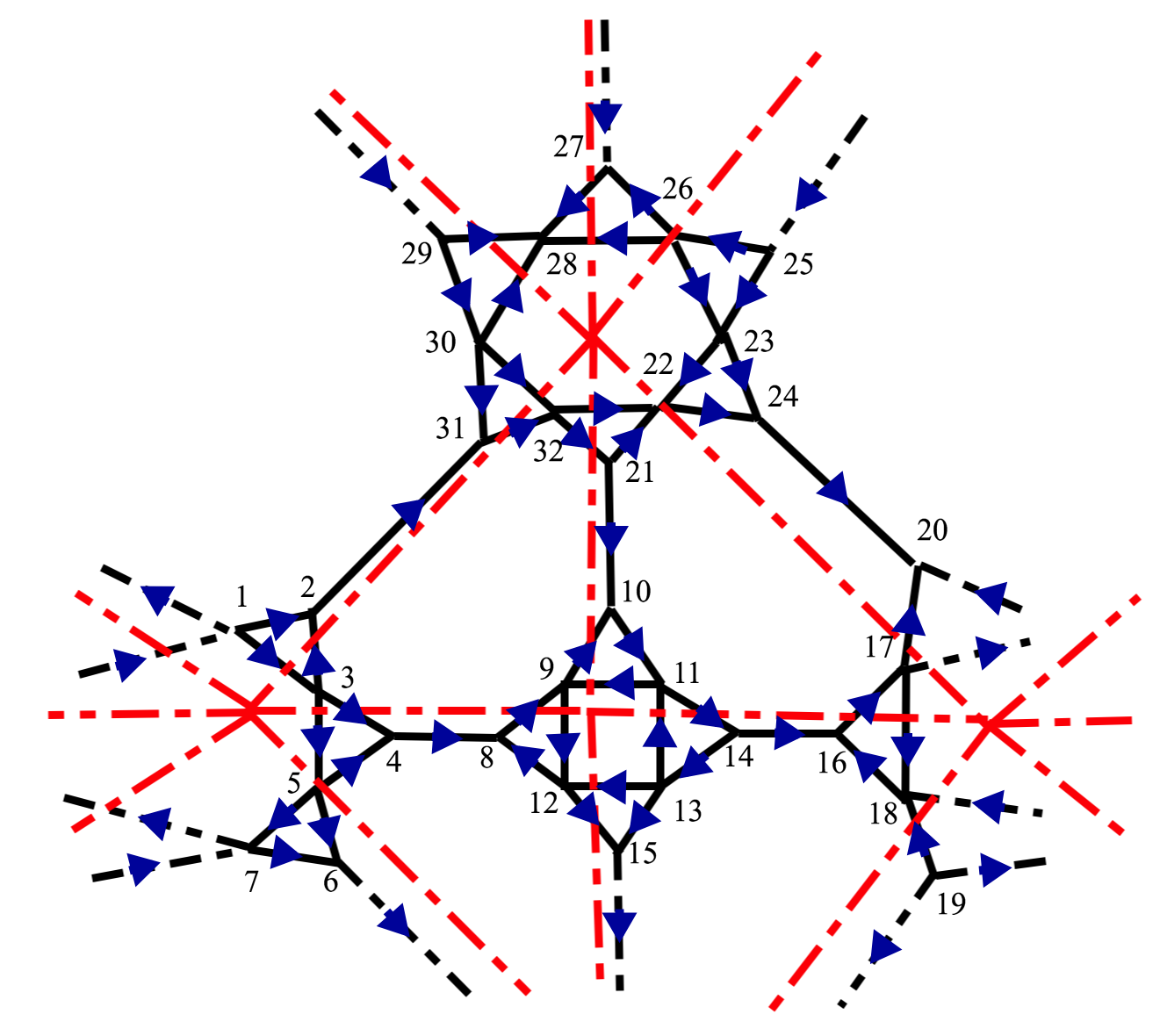}
    \caption{Fisher-Dubed\'{a}t decoration for the staggered Ising model lattice.}
    \label{fig:fddec}
\end{figure}

Of course, once the mapping is found, we still need to find the solution of the dimer problem. This can be done for translational invariant graphs, and here we assume that the  graph has toroidal boundary conditions. We sketch the steps here for completeness and to explain the notation. If $\tilde A$ is the oriented and weighted graph of the Fisher-Dub\'{e}dat construction, and if the graph is translational invariant, then the total number of dimer configurations can be calculated given via $Z_{dimer}=\frac{1}{2(2\pi)^2} \int_0^{2\pi}  \int_0^{2\pi}   d\theta d\phi \log \text{det}(A(\phi,\theta))$, where $A(\phi,\theta)$ is calculated by identifying a repeatable motif on the graph. The motif is shown in Fig. \ref{fig:dubedat}, and we assume that the graph has $m$ and $n$ vertical and horizontal repetitions of same pattern.
As it is known, this implies that the matrix $A$ can be arranged in blocks, parametrized by two integer numbers which represent the symmetry of the rectangular lattice, such that $A_{ij}\equiv A_{(x,y),(x^\prime,y^\prime)}$.  Then, we construct the following matrix:
\begin{equation}
    A(\theta,\phi)=A_0+A_y^{+}e^{i\theta} +A_y^{-}e^{-i\theta}+A_x^{+}e^{i\phi}+A_x^{-}e^{-i\phi}
\end{equation}
and in the limit in which the $mn\rightarrow \infty$, it is possible to show that
\begin{equation}
\lim_{mn\rightarrow \infty}   \frac{\log \text{det}(\tilde A)}{mn} =\int_{0}^{2\pi}\int_{0}^{2\pi} d\theta d\phi \log \text{det} \lambda(\theta,\phi).
\end{equation}
This implies that the free energy \textit{per site} can be actually calculated via the determinant on the smaller motif of the system. The matrix $A_0$ represent connections inside the motif, while $A_y^+$ and $A_y^-$ are connections of the same motif to the motif above and below respectively, and $A_x^+$ and $A_x^-$ are connections to the right and left respectively. This construction is well documented and the details thus omitted, these steps should clarify the derivation. This said, evaluating the determinant $\lambda(\phi,\theta)$ can be challenging. For the case of Fig. \ref{fig:dubedat}, the matrix $\lambda(\phi,\theta)$ is a sparse 32$\times$ 32 matrix, as the minimum number of nodes of the motif is 32. The decoration of the lattice is in fact shown in Fig. \ref{fig:fddec}. 

After some calculations, we find that if we define
%\begin{widetext}
\begin{flalign}
\footnotesize
\lambda(\phi,\theta)=\left(
\begin{array}{cccccccc}
 0 & 1 & -1 & -1 & 0 & z & 0 & 0 \\
 -1 & 0 & 1 & -1 & -z e^{-i \theta} & 0 & 0 & 0 \\
 1 & -1 & 0 & 1+z e^{i \phi} & 0 & 0 & 0 & 0 \\
 1 & 1 & -1-z e^{-i \phi} & 0 & 0 & 0 & 0 & 0 \\
 0 & z e^{i \theta} & 0 & 0 & 0 & 1 & -1 & -1 \\
 -z & 0 & 0 & 0 & -1 & 0 & 1 & -1 \\
 0 & 0 & 0 & 0 & 1 & -1 & 0 & 1+e^{i \phi} z_s \\
 0 & 0 & 0 & 0 & 1 & 1 & -1-e^{-i \phi} z_s & 0 \\
\end{array}
\right)
\end{flalign}
%\end{widetext}
where $z=\tanh(\frac{J}{\kappa T})$ and $z_s=\tanh((2s-1)\frac{J}{\kappa T})$,  where $\kappa$ is the Boltzmann constant, and the partition function of the model is given by
\begin{eqnarray}
f=\frac{\log (Z)}{N}=\log 2+\frac{1}{4(2\pi)^2} \int_0^{2\pi} d\phi\int_0^{2\pi} d\theta \log\Big( \mathcal C(s,T)\text{det}(\lambda(\phi,\theta)\big)\Big)
\label{eq:partfuncext}
\end{eqnarray}
where $\mathcal C(s,T)=\Big(\cosh(\frac{J}{\kappa T})^3\cosh((2s-1)\frac{J}{\kappa T})\Big)^2$.


Following the procedure above, we obtain the solution can be written in the following form
\begin{eqnarray}
    f=g(\beta)+\frac{1}{2 (2\pi)^2} \int_0^{2\pi} d\theta\int_0^{2\pi} d\phi \log \text{det}(\lambda(\nu,\nu_d,\phi,\theta)\nonumber 
\end{eqnarray}
where $\lambda$ is the matrix obtained in the previous section,  and $g(\beta)=\log \prod \cosh(v_d) \prod \cosh(v) $, with $\beta=1/(\kappa T)$.

Obtaining the determinant analytically is a hard task for matrices of this size by hand. However, we employed an algebra software. 
The determinant can be written as
\begin{eqnarray}
    \text{det}\ \lambda&=&a(\nu,\nu_d)+b(\nu,\nu_d) \cos (\theta )+c(\nu,\nu_d) \cos (\phi ) \nonumber \\
    & &+d(\nu,\nu_d) \cos (\theta -\phi )+e(\nu,\nu_d) \cos (\theta +\phi ),\nonumber 
\end{eqnarray}
where the functions $a...e$ are reported in eqn. (\ref{eq:sol}) as a function of the parameters $v$ and $v_d$.

\begin{flalign}
%det M&=& a+b \cos (\theta )+c \cos (\phi )+d \cos (\theta -\phi )+e \cos (\theta +\phi ) \\
   a(\nu,\nu_d)&=-16 \Big(-3 \nu ^8 \nu_d^4-\nu^8+4 \nu^6 \nu_d^4+16 \nu^6 \nu_d^3+4 \nu^6 \nu_d^2-10 \nu^4 \nu_d^6\nonumber \\
   &\ \ \ \ \ \ \ \  -12 \nu^4 \nu_d^5+13 \nu^4 \nu_d^4+8 \nu^4 \nu_d^3-2 \nu^4 \nu_d^2 -12
   \nu^4 \nu_d-5 \nu^4 \nonumber \\
   &\ \ \ \ \ \ \ \  -12 \nu^2 \nu_d^5-8 \nu^2 \nu_d^4-8 \nu^2 \nu_d^3-8 \nu^2 \nu_d^2-12 \nu^2 v_d-12 \nu_d^4-4\Big) \nonumber \\
   b(\nu,\nu_d)&=-16 \Big(-2 \nu^8 \nu_d^4+2 \nu^8 \nu_d^2-8 \nu^6 \nu_d^5-2 \nu^6 \nu_d^4+8 \nu^6 \nu_d^3+2 \nu^6+6 \nu^4 \nu_d^6+12 \nu^4 \nu_d^5 \nonumber \\
   &\ \ \ \  \ \ \ \  +18 \nu^4 \nu_d^4 -24 \nu^4 \nu_d^2-12
   \nu^4 v_d+20 \nu^2 \nu_d^5+8 \nu^2 \nu_d^4 \nonumber \\
   &\ \ \ \ \ \ \ \  \ \ \ \   -8 \nu^2 \nu_d^3-12 \nu^2 v_d-8 \nu^2+8 \nu_d^4-8 \nu_d^2\Big) \nonumber \\
   c(\nu,\nu_d)&= -16 \Big(2 \nu^8 \nu_d^4-2 \nu^8 \nu_d^2+2 \nu^6 \nu_d^5-4 \nu^6 \nu_d^4-8 \nu^6 \nu_d^3+6 \nu^6 v_d+4 \nu^6+6 \nu^4 \nu_d^5+6 \nu^4 \nu_d^4 \nonumber \\
   &\ \ \ \  \ \ \ \   -6 \nu^4 \nu_d^2-6 \nu^4
   v_d-8 \nu^2 \nu_d^5+4 \nu^2 \nu_d^4+8 \nu^2 \nu_d^3-4 \nu^2-8 \nu_d^4+8 \nu_d^2\Big)\nonumber \\
   d(\nu,\nu_d)&=  -16 \Big(4 \nu^6 \nu_d^5-8 \nu^6 \nu_d^3+4 \nu^6 v_d-8 \nu^4 \nu_d^5 \nonumber \\
   &\ \ \ \ \ \ \ \   +16 \nu^4 \nu_d^3-8 \nu^4 v_d+4 \nu^2 \nu_d^5-8 \nu^2 \nu_d^3+4 \nu^2 v_d\Big) \nonumber \\
   e(\nu,\nu_d)&=-16 \Big(2 \nu^8 \nu_d^4-2 \nu^8 \nu_d^2-2 \nu^6 \nu_d^5+2 \nu^6 v_d-2 \nu^4 \nu_d^5 \nonumber \\
   & \ \ \ \ \ \ \ \  -10 \nu^4 \nu_d^4+10 \nu^4 \nu_d^2+2 \nu^4 v_d+4 \nu^2 \nu_d^5-4 \nu^2
   \nu_d+8 \nu_d^4-8 \nu_d^2\Big).\label{eq:sol} \\
  \nu_d&=\tanh(v_d), \nu=\tanh(v).\nonumber 
   \label{eq:sol}
\end{flalign}

\begin{figure}
    \centering
    \includegraphics[scale=0.4]{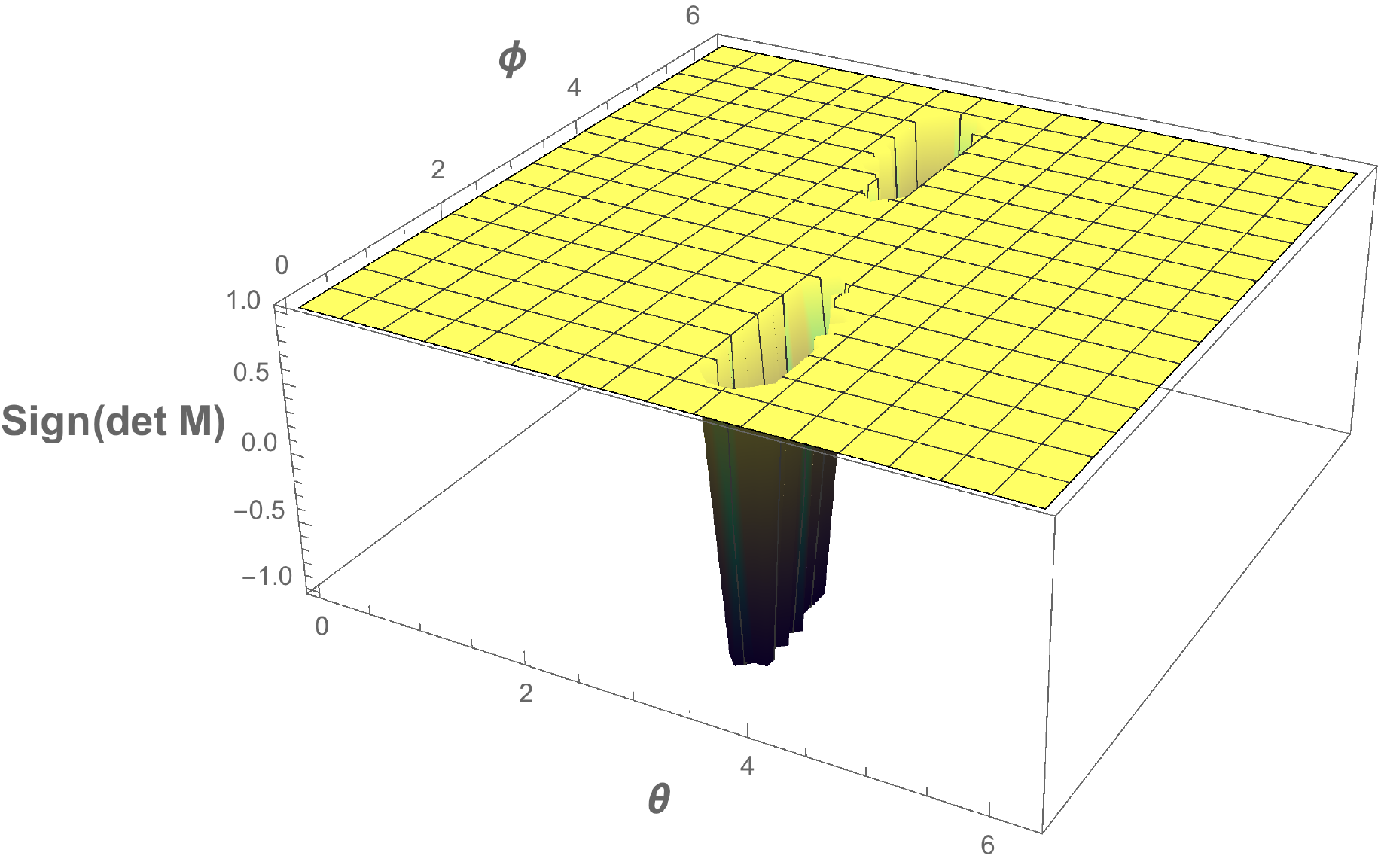}
    \caption{Non-analyticity of the function $M(v,v_d,\theta,\phi)$ as a function of $\theta$ and $\phi$ for $v=3$ and $v_d=4$. We see that in the interval of integration for the partition function the function $\lambda$ can become negative, signalling a non-analyticity of the partition function.}
    \label{fig:neg}
\end{figure}

First, we note that the model presents a phase transition, due to the fact that $\lambda(v,v_d,\phi,\theta)$ can be negative, as shown in Fig. \ref{fig:neg}, but is otherwise non-trivial to characterize analytically. This implies in turn a non-analyticity of the partition function in the set of the two parameters $v$ and $v_d$ for this particular type of vertex model.
On the other hand, if we use the monopole gas parametrization, it is not hard to see numerically that $det\ \lambda>0$ always, which implies that this model does not exhibit any phase transition.

\section{The Onsager-Villain model: tunable frustration}

\begin{figure}
    \centering
    \includegraphics[scale=1]{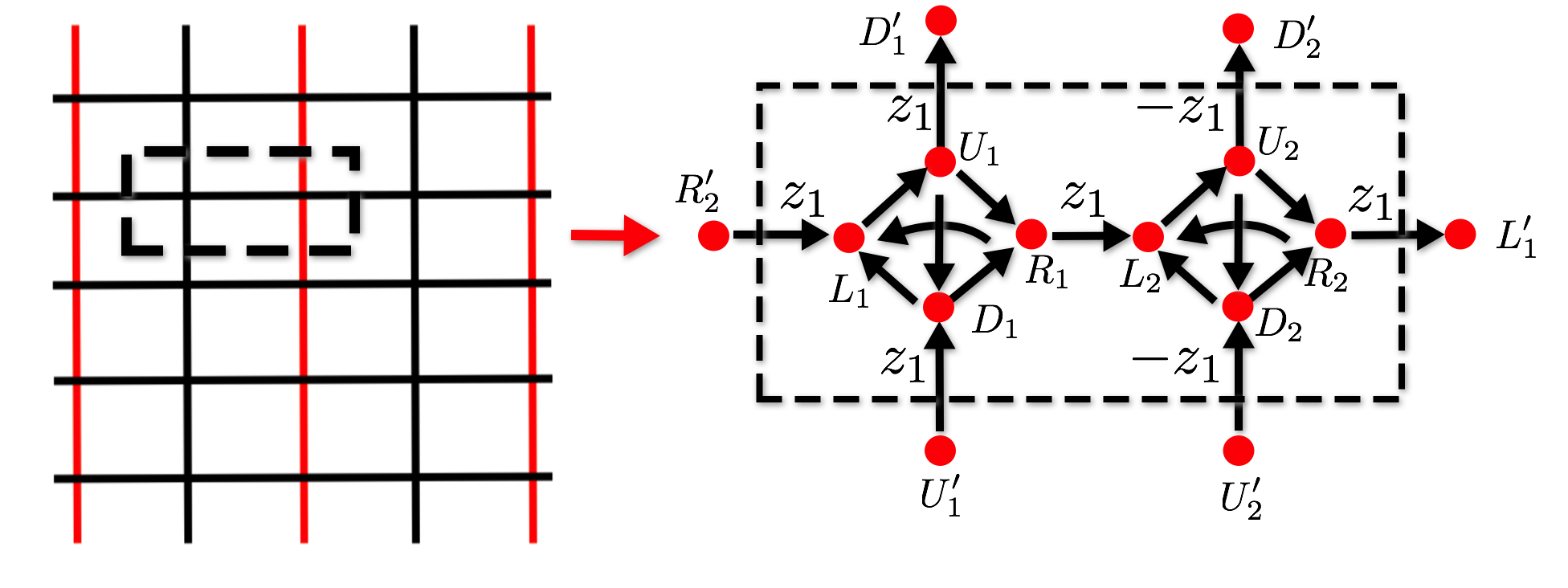}
    \caption{The transformation from the Villain's model to the MPW dimer model. Black are ferromagnetic interactions while red are antiferromagnetic interactions.}
    \label{fig:villain}
\end{figure}

Another exactly solvable model with tunable frustration is the following, based on an earlier model introduced by Villain \cite{villain}, also called Villain's fully frustrated model. This is an exactly solvable spin model on a square lattice, in which at every two columns (or rows) we have antiferromagnetic couplings. The solution can be obtained by dimer techniques, and is given by
\begin{eqnarray}
   \frac{\log Z}{N}&=&\log (2\cosh(2J/\kappa T))\nonumber \\
   &&+\frac{1}{4\pi^2} \int_0^\pi d\phi \int_0^\pi d\theta \log\Big( (1+\tilde z^2)^2-2\tilde z^2 \cos \phi-2\tilde z^2 \cos \theta\Big)
\end{eqnarray}
with $\tilde z = \tanh \big(2J/\kappa T\big)$. The expression is obtained using the Montroll-Potts-Ward technique \cite{montroll}, which is a decoration of the type shown in Fig. \ref{fig:villain}.
Then, one calculates the solution of the dimer problem using standard techniques, which involves the evaluation of a determinant analytically, as done in the previous section.

Instead, we consider the following Hamiltonian
\begin{equation}
    H=(1-s) H_{Ising-2D}+s H_{Villain},
    \label{eq:oddhd}
\end{equation}
where $H_{Ising-2D}$ is the Hamiltonian of the 2D and ferromagnetic Ising  model without external field, while $H_{Villain}$ is the Villain's fully frustrated Ising model \cite{villain}. For this reason, we call this model the Onsager-Vilalin model. If we call the ferromagnetic couplings for the Ising and Villain's model $-J$, and the antiferromagnetic couplings of Villain's model $+J$, then the model has the most spins with ferromagnetic couplings invariant in value, while the every two columns the couplings become $(2s-1)J$. Thus, we can solve the model using the same technique used by Villain which is using the Montroll-Potts-Ward dimer model mapping. 

The techniques are standard and can be found in \cite{montroll}, and as in the case of \cite{villain}, we end up with $8\times 8$ determinant of a matrix $\lambda$.
After some calculations, we find that for the Onsager-Villain's model the matrix $\lambda$ is given by:
%\begin{widetext}
\begin{eqnarray}
\footnotesize
\lambda(\phi,\theta)=\left(
\begin{array}{cccccccc}
 0 & 1 & -1 & -1 & 0 & z & 0 & 0 \\
 -1 & 0 & 1 & -1 & z \left(-e^{-i \theta}\right) & 0 & 0 & 0 \\
 1 & -1 & 0 & 1+z e^{i \phi} & 0 & 0 & 0 & 0 \\
 1 & 1 & -1+z \left(-e^{-i \phi}\right) & 0 & 0 & 0 & 0 & 0 \\
 0 & z e^{i \theta} & 0 & 0 & 0 & 1 & -1 & -1 \\
 -z & 0 & 0 & 0 & -1 & 0 & 1 & -1 \\
 0 & 0 & 0 & 0 & 1 & -1 & 0 & 1+e^{i \phi} z_s \\
 0 & 0 & 0 & 0 & 1 & 1 & -1-e^{-i \phi} z_s & 0 \\
\end{array}
\right)
\end{eqnarray}
%\end{widetext}
where $z=\tanh\big(\frac{J}{\kappa T}\big)$ and $z_s=\tanh\big((2s-1)\frac{J}{\kappa T}\big)$,  the partition function of the model is given by
\begin{eqnarray}
\frac{\log (Z)}{N}=\log 2+\frac{1}{4(2\pi)^2} \int_0^{2\pi} d\phi\int_0^{2\pi} d\theta \log\Big( \mathcal C(s,T)\text{det}(\lambda(\phi,\theta)\big)\Big)
\label{eq:partfuncext}
\end{eqnarray}
where $\mathcal C(s,T)=\Big(\cosh\big(\frac{J}{\kappa T}\big)^3\cosh\big((2s-1)\frac{J}{\kappa T}\big)\Big)^2$.

We can evaluate the determinant exactly, to 
\begin{flalign}
\det \lambda_s(\phi,\theta)&=z_s \Big(z_s \left(z^6+2 \left(z^4-1\right) z \cos \left(\phi\right)+z^4-2 \left(z^2-1\right) z^2 \cos \left(\phi
   _2\right)+z^2+1\right)\nonumber\\
   &  \ \ \ \ \ \ +2 z \left(z^2-1\right)^2 \cos \left(2 \phi\right)+2 \left(z^2+1\right)^2 \left(\left(z^2-1\right) \cos
   \left(\phi\right)+z\right)\Big) \nonumber \\
   &+2 z \left(z^2-1\right) \left(\left(z^2+1\right) \cos \left(\phi\right)+z \cos \left(\phi
   _2\right)\right)+\left(z^2+1\right) \left(z^4+1\right)
\end{flalign}
For $z_s\rightarrow-z$, we have
\begin{eqnarray}
\det \lambda_{s=1}(\phi,\theta)&=&\left(z^2-1\right)^2 \Big(\left(z^2+1\right)^2\nonumber \\
&-&2 z^2 \left(\cos \left(2 \phi\right)+\cos \left(\theta\right)\right)\Big)
\end{eqnarray}
which is Villain's expression provided we rescale the temperature. For $z_s\rightarrow z$ we have
\begin{eqnarray}
\det \lambda_{s=0}(\phi,\theta)&=&2 z \left(z^2-1\right) \Big(2 \left(z^2+1\right)^2 \cos \left(\phi\right)\nonumber \\
&+&z \left(z^2-1\right) \left(\cos \left(2 \phi\right)-\cos
   \left(\theta\right)\right)\Big)+\left(z^2+1\right)^4
\end{eqnarray}
If we perform a high temperature expansion of the expression
\begin{flalign}
&\log \left(\frac{2 z \left(z^2-1\right) \left(\cos \left(\phi\right)+\cos \left(\theta\right)\right)+\left(z^2+1\right)^2}{\sqrt{2 z
   \left(z^2-1\right) \left(2 \left(z^2+1\right)^2 \cos \left(\phi\right)+z \left(z^2-1\right) \left(\cos \left(2 \phi\right)-\cos
   \left(\theta\right)\right)\right)+\left(z^2+1\right)^4}}\right)\nonumber \\
   &=\sum_k f_k(\phi,\theta) z^k\nonumber
\end{flalign}
where the first terms are 

\begin{flalign}
f_1(\phi,\theta)&=2  \cos \left(\theta\right)\nonumber \\
f_2(\phi,\theta)&=\sin ^2\left(\theta\right)-\cos \left(\theta\right) \left(4 \cos \left(\phi\right)+\cos
   \left(\theta\right)-1\right)\nonumber \\
f_3(\phi,\theta)&= 4 \cos ^3\left(\phi\right)-\frac{8}{3} \cos ^3\left(\theta\right)-8 \cos \left(\phi
   _2\right) \cos ^2\left(\phi\right)+6 \cos \left(\theta\right) \nonumber \\
   & +4 \cos \left(\phi\right) \left(\sin ^2\left(\phi\right)-2 \cos
   ^2\left(\theta\right)+\cos \left(\theta\right)\right)\nonumber \\
f_4(\phi,\theta)&= \left(6 \cos \left(2 \phi\right)-4 \cos \left(3 \phi
   _1\right)+2\right) \cos \left(\theta\right)-\frac{3}{2} \left(4 \cos \left(2 \phi\right)+1\right) \cos \left(2 \theta\right) \nonumber \\
   &\ \ \ \ \  -4 \cos
   \left(\phi\right) \cos \left(3 \theta\right)-\frac{1}{2} \cos \left(4 \theta\right)\nonumber 
\end{flalign}
we note that our result seem to differ from the partition function of the 2D Ising model.
Clearly our is not Onsager's expression, and thus one may deem such limit wrong. However, one can promptly see that 
\begin{eqnarray}
\frac{1}{(2\pi)^2} \int_0^{2\pi}d\phi\int_0^{2\pi}d\theta f_k(\phi,\theta)=0
\end{eqnarray}
by a direct calculation.

Thus, the expression we obtained can be replaced inside the integral as
\begin{eqnarray}
\log(\det \lambda_{s=0})=2\log\left(2 z \left(z^2-1\right) \left(\cos \left(\phi\right)+\cos \left(\theta\right)\right)+\left(z^2+1\right)^2\right)
\end{eqnarray}
and finally we obtain that eqn. (\ref{eq:partfuncext}) can be reduced to Onsager's solution as expected.

\subsection{Analytical properties of the model}
We now ask ourselves whether the model is analytical, by looking at points in which $\det F<0$. In order to ask this question, we set $\phi=\theta=0$, and look for a relationship between $z_s$ and $z$ by setting $\det \lambda=0$.

A solution can be obtained, given by
\begin{eqnarray}
   z_s=-\frac{-z^3-z^2-z+1}{z^3+z^2-z+1}, 
\end{eqnarray}
or, for $J=1$,
\begin{eqnarray}
   s(T)=\frac{1}{2} \left(T \tanh ^{-1}\left(\frac{z^3+z^2+z-1}{z^3+z^2-z+1}\right)+1\right)
\end{eqnarray}
It is interesting to look at the points in which $s(T)=0$.
This is given by the solution of
\begin{eqnarray}
   \frac{z^3+z^2+z-1}{z^3+z^2-z+1}+z=0,
\end{eqnarray}
which is given by
\begin{eqnarray}
   z=\tanh\frac{J}{T}=\sqrt{2}-1.
\end{eqnarray}
It is not hard to see that the solution is exactly the critical temperature of the $2D$ Ising model, $T_c=2.26919[..] J$.
\begin{figure}
    \centering
    \includegraphics[scale=1.5]{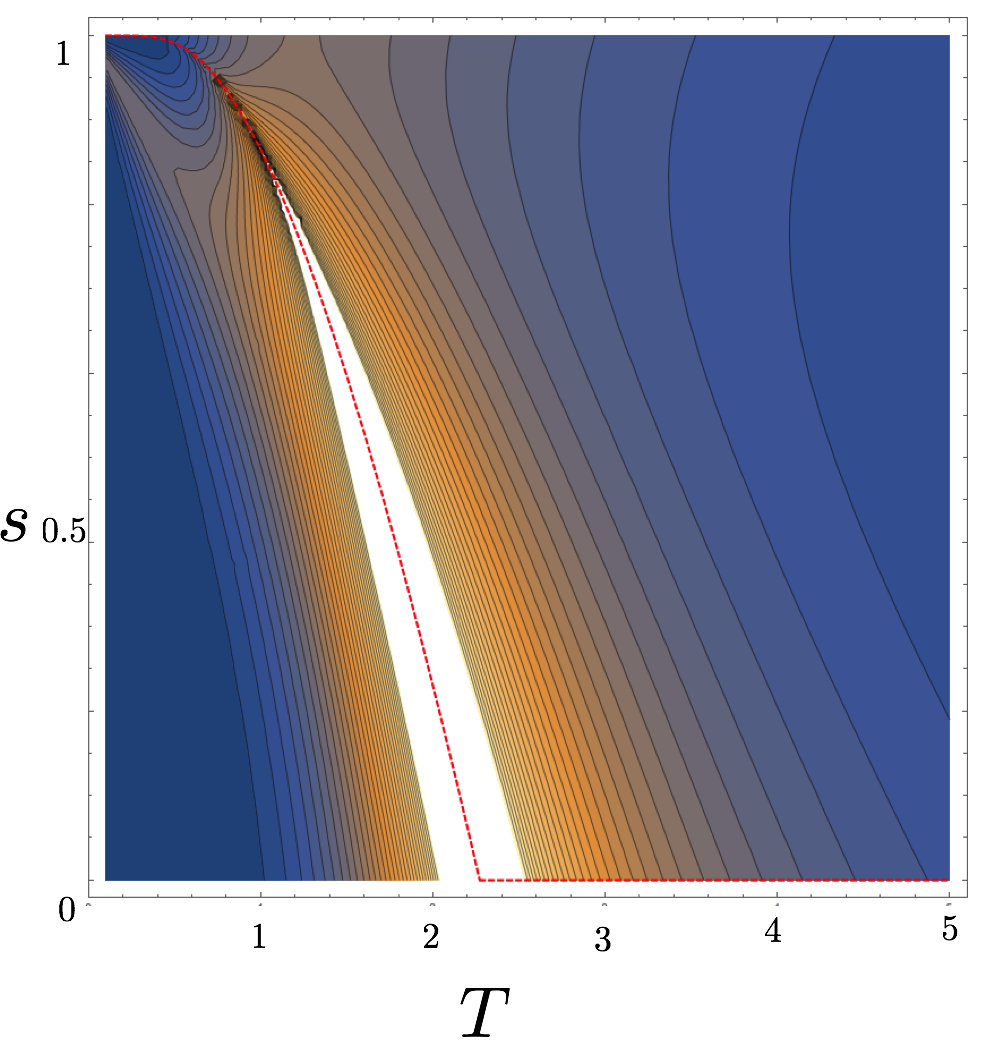}
    \caption{Contour map of the heat capacity of the Onsager-Villain model as a function of $T$ and s. The dashed line represents a critical line evaluated analytically. It is easy to see that the ordered phase and the disordered phase of the model are not the same, being one not able to continuously go by analytical continuation in the $s$ direction, from the disordered high temperature phase to the ordered low temperature phase of the Ising model.}
    \label{fig:ovcv}
\end{figure}

\section{Spin ice Bethe approximation}
The last problem we wish to address is the problem of evaluating the number of spin ice configurations on a tree graph, a problem of recent experimental interest \cite{sacconetree}.

Let us consider a spin ice problem of the form
\begin{eqnarray}
   H=\sum_{v\in \mathcal G} S_v(\{s_{i\rightarrow v}\}).
   \label{eq:edges}
\end{eqnarray}
We assume that $\mathcal G=(V,E)$ is a graph whose edges label the spin, and 
where $G_v$ is a function which evaluates the energy at each vertex. 
A spin ice Hamiltonian is such that $s_i=\pm 1$ depending on the orintation, and $G_v\geq 0$, while $G_v(\{s_{i\in v}\})=0$ if $\sum_{i\in v} s_i=0$. A balanced graph is such that all nodes are in the spin ice state. An example of a function satisfying the properties above is given by
\begin{eqnarray}
   S_v=(\sum_{i} B_{vi} s_i)^2.
\end{eqnarray}
where $B_{vi}$ is the directed incidence matrix of the graph $\mathcal G$.
\begin{figure}
    \centering
    \includegraphics[scale=0.5]{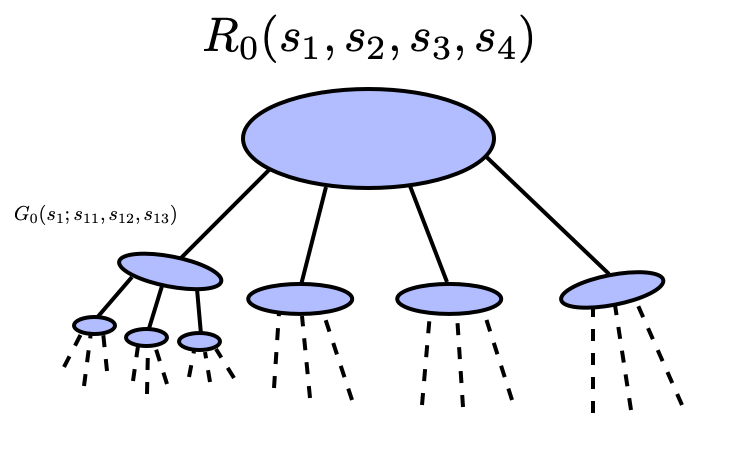}
    \caption{The key idea behind the Bethe equation. Each vertex node but the root is converted into a gate. Starting from the leaves, then one to integrate out progressively the spins until the root is reached.}
    \label{fig:spinice}
\end{figure}
For the purpose of this section we consider graphs $\mathcal G$ which are trees of even coordination. The partition function of the problem can be written as
\begin{eqnarray}
   Z=\sum_{s_i=\pm 1} e^{-\beta H}=\sum_{s_i=\pm 1}\prod_{v} e^{-\beta S_v}.
\end{eqnarray}
Given that our graph is a tree, we can identify a boundary $\mathcal B$ given by the nodes of degree $1$ which we call leaves. Obviously, we can also define a root of the tree, which we call $v_0$.
It is not hard to see at this point why the Bethe method can be employed to solve exactly this model. Let us identify the leaf vertices $\mathcal S_i\in \mathcal B$. We call $\mathcal B_v$ all the nodes at distance $1$ from the boundary.
Then, we can iteratively calculate the partition function starting from the leaves, and working our way to the root vertex.

Since the graph is a tree, the vertices $B_v$ must not have any edge in common, or the graph would contain loops, and also all the leaves can be grouped in a way to be connected to the vertices of $B_v$. We can then partition the set of leaves in
\begin{eqnarray}
\mathcal B&=&\{ \{B_v^1\},\{B_v^2\},\cdots \} \nonumber \\
&=&\{\{ s_1,\cdots, s_{m_1}\},\{ s_{m_1+1},\cdots, s_{m_1+m_2}\},\cdots \},
\end{eqnarray} where $B_v^1$ represents the edges in the first set in $\mathcal B_v$ and so on.
The integration over the leaves can be done as $\sum_{\mathcal S\in B_v^1}e^{-\beta G_{v_1}}\sum_{\mathcal S\in B_v^2}e^{-\beta G_{v_2}}\cdots $. Since $G_{v_1}$ is the vertex energy of a rooted tree, it will depend on $m_1+1$ edges, where the $+1$ is the edge connected to the higher vertex in the hierarchy, as in Fig. \ref{fig:spinice}. %One way to express this is to write the spins in tree coordinates. If the root has coordination $k_0$, then we have $k_0$ branches. A tree edge coordinate is then
%\begin{eqnarray}
%  &&\{\{s_1,\cdots, s_{k_0}\},\{s_{11},\dots,s_{1,k_1^1}\},\\
%  &&\{s_{21},\dots,s_{1,k_{21}^1}\},\cdots,\{s_{111},s_{112},\cdots,s_{11k_{11}^1}\},\cdots,\},\nonumber 
%\end{eqnarray}
%where $s_{abcd\cdots}$ is the spin with coordinate ${abcd\cdots}$. The length of the coordinate represents the layer at which the spins are. However, since the graph is a rooted tree, we can identify up to the root the edge with the vertex below it.
We can write then
\begin{eqnarray}
   \sum_{\{s\in \mathcal B_v\}=\pm 1}^{d_v-1} e^{-\beta H_{v}(s_v,\{s\in \mathcal B_v\})}=F_{v}(s_v)
\end{eqnarray}
We can call $e^{-\beta S_{v}(s_v,\{\mathcal S\in \mathcal B_v\})}=G_0(s_v;\{s\in \mathcal B_v\})$ and $e^{-\beta S_{0}(s_v,\{s\in \mathcal B_v\})}=R_0(s_1,\cdots,s_k)$.
We see immediately that the algorithm leads to a message passing, and thus is a form of Bethe approximation. At zero temperature, $F_v$ is simply the number of configurations which are compatible with the value of $s_v$ leaving the node. For instance, if the graph has coordination $4$, there are $3$ spin configurations compatible with the exiting value $s_1=\pm 1$. It turns out that $F_{v}(s_v)\equiv F_0=3$.
To see why this algorithm is powerful, consider a rooted graph of coordination $4$, and of depth $L$. Then, immediately we can write
\begin{eqnarray}
   Z&=&\sum_{s_1,\cdots,s_d=\pm 1} R_0(s_1,\cdots,s_d)F_{d-1}^{\frac{(d-1)^{L+1}-1}{d-2}}\\
   &=&C_d F_{d-1}^{\frac{(d-1)^{L+1}-1}{d-2}},
\end{eqnarray}
where $C_d=\sum_{s_1,\cdots,s_d=\pm 1} R_0(s_1,\cdots,s_d)$ is the number of spin ice configurations allowed by a vertex of coordination $d$, while $F_{d-1}$ is the number of configurations compatible with a spin up and down, e.g. with sum equal to $\pm 1$.
Let us set $d=4$, for which $C_d=6$ and $F_{d-1}=3$; we have then
\begin{eqnarray}
   Z=\frac{3}{4} 3^{2(3^{L+1}-1)}
   \label{eq:Spinice4}
\end{eqnarray}
The number of nodes in the graph for $d=4$, is $N_v=6 \left(3^L-1\right)+1$. It follows that the entropy per node, as $L\rightarrow \infty$, is then 
\begin{eqnarray}
   s=\lim_{L\rightarrow \infty}\frac{\log Z}{N_v}=\log 3,
\end{eqnarray}
from which we can see that the model is frustrated.

\begin{figure}
    \centering
    \includegraphics[scale=0.45]{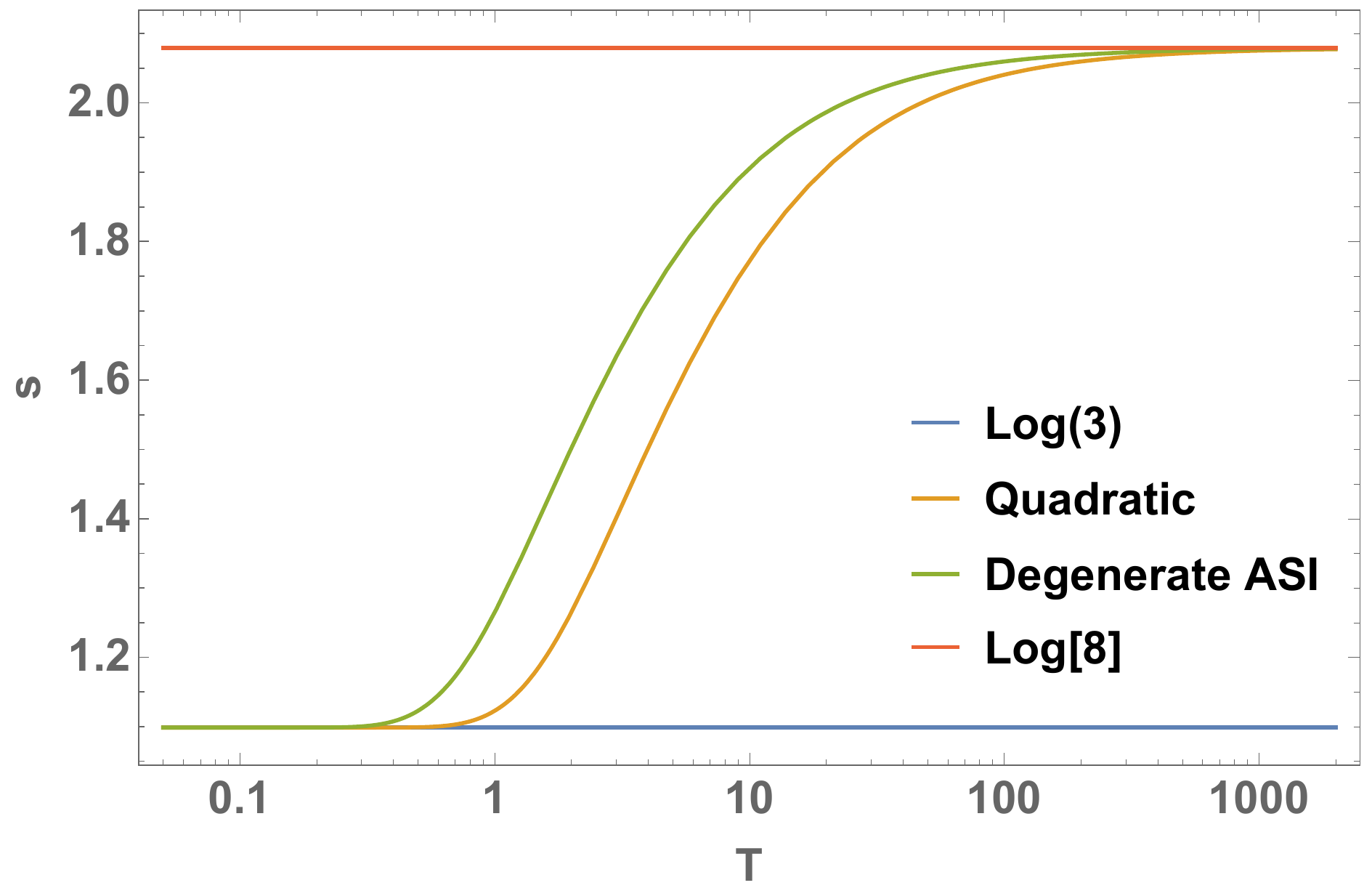}
    \caption{Entropy per node at finite temperature divided as a function of the temperature for the degenerate ASI model and the quadratic model. }
    \label{fig:entroppnt}
\end{figure}

For the case at finite temperature, we have for $d=4$ and $\beta$ finite,
\begin{eqnarray}
   F_{d-1}&=&e^{-16 \beta }+4 e^{-4 \beta }+3,\\
    C_d&=&2 e^{-16 \beta }+8 e^{-4 \beta }+6.
\end{eqnarray}
We can then plot the entropy as a function of the temperature.
The result of the entropy per node is shown in Fig. \ref{fig:entroppnt}.

At high temperature, $C_d\approx 16$ while $F_{d-1}\approx 8$ and in the limit $L\rightarrow \infty$ the entropy per note becomes $\log 8$, which are completely uncorrelated nodes. The results thus interpolates between a minimum and a maximum entropy per node allowed.

Another effective model is given by
\begin{eqnarray}
   G_v=e^{\beta  (-E_{\|} (s_1 s_3+s_2 s_4+2)-E_{\perp}
   ((s_1-s_3) (s_2-s_4)+4))}
\end{eqnarray}
is commonly used in artificial square ice (ASI).
For the square spin ice, it has been noted that vertices have four increasing energies parametrized by $\epsilon_{\perp}$ and $\epsilon_{||}$, with a nomenclature Type I,$\cdots$,Type IV respectively. The vertex energies are
    $\epsilon_{I}=-4 \epsilon_{\perp}+2 \epsilon_{||}$,  $\epsilon_{II}=-2 \epsilon_{||}$, \\
    $\epsilon_{III}=0$,
    $\epsilon_{IV}= 4\epsilon_{\perp}+2\epsilon_{||}$,
where $\epsilon_{I}<\epsilon_{II}<\epsilon_{III}<\epsilon_{IV}$. The vertex population in the ground state is determined by this energy hierarchy. In units of the temperature for $\kappa=1$, we can use 
$\epsilon_{\perp}\approx0.38675$ and $\epsilon_{\|}\approx 0.2735$ for realistic phase diagrams, as noted in \cite{Morrison}. Here we are interested in the degenerate case, $E_{\perp}=E_{\|}=1$.
We obtain for $d=4$:
\begin{eqnarray}
F_{3}&=&e^{-8 \beta }+4 e^{-2 \beta }+3 \nonumber  \\
C_4&=&2 e^{-8 \beta }+8 e^{-2 \beta }+6.\nonumber
\end{eqnarray}
We have shown a comparison between the quadratic and degenerate ASI model in Fig. \ref{fig:entroppnt}. Both curves interpolate between $\log(3)$ and $\log(8)$, but the difference between the two is in how the energetics of the monopoles. A comment is that in the typical Pauling calculation the entropy per node is half of what we evaluated; however in the tree calculation it is notorious that the bulk is as large as the boundary, and thus the entropy per node is twice as much.

As a last example, consider a spin ice tree of coordination $3$ in the spin ice ground state, in which only vertices with monopoles of charge $\pm1$ are allowed. One can think of this tree as a Kagome lattice in which loops have been removed in a symmetrical way. It is not hard to perform this counting, and we see that also in this case, $F_2=3$ and $C_3=6$, exactly as in the case of coordination four; it follows that the same result of eqn. (\ref{eq:Spinice4}) applies, provided that the coordination is changed from four to three.

\section{Conclusions} Geometric frustration is an active area of research. There is only a limited number of non-trivial frustrated models that can be solved exactly
\cite{Levis1,Levis2,Levis3}. Ice models, e.g. the Rys-F, the 6- and 8- vertex models are the typical examples of exactly solvable models whose solution can be obtained via the Bethe Ansatz, using a line formalism \cite{Baxter}. Typically, these models are complicated and solutions are hard to be obtained via standard techniques. In this paper we studied some models that can be solved exactly, with different degrees of complexity. In the literature, various exact methods for the solutions of frustrated spin systems \cite{vuoto1,vuoto2,vuoto3,vuoto5,vuoto5}, including those with disorder and employing the Bethe ansatz \cite{disorder1,disorder2,disorder3}, including the Bethe Permanent \cite{disorder4}. This paper contributes to this literature, and in future papers we will discuss more applications of these exactly solvable models.

\vspace{1cm}

\textbf{Acknowledgements.} This work was carried out under the auspices of the NNSA of the U.S. DoE at LANL under Contract No. DE-AC52-06NA25396. FC was also financed via DOE-ER grant PRD20190195. We thank W. Cunningham and Y. Subasi for allowing us to publish the solution of the Onsager-Villain model separately from our joint paper.
\vspace{1cm}

\end{document}